\documentclass{aip-cp}

\usepackage[numbers]{natbib}
\usepackage{rotating}
\usepackage{graphicx}

\usepackage{combelow}
\usepackage{bm}
\usepackage{slashed}
\usepackage{braket}
\usepackage{color}

\def\eqref#1{(\ref{#1})}

\def\halpha{{\hat{\alpha}}}
\def\hbeta{{\hat{\beta}}}

\def\htau{{\hat{\tau}}}
\def\heta{{\hat{\eta}}}

\def\hatx{{\hat{x}}}
\def\haty{{\hat{y}}}
\def\feq{\ensuremath{f^{\rm{(eq)}}}}
\def\etas{(\overline{\eta / s})}

\begin{document}

\title{Lattice Boltzmann study of the one-dimensional boost-invariant expansion 
with anisotropic initial conditions}

\author[aff1]{Victor E. Ambru\cb{s}\corref{cor1}}

\author[aff1]{C\u{a}lin Guga-Ro\cb{s}ian\corref{cor2}}

\affil[aff1]{Department of Physics, West University of Timi\cb{s}oara,
	Bd.~Vasile P\^arvan 4, Timi\cb{s}oara 300223, Romania}
\corresp[cor1]{Corresponding author: Victor.Ambrus@e-uvt.ro}
\corresp[cor2]{calinguga@gmail.com}

\maketitle

\begin{abstract}
A numerical algorithm for the implementation of anisotropic distributions 
in the frame of the relativistic Boltzmann equation is presented. 
The implementation relies on the expansion of the Romatschke-Strickland 
distribution with respect to orthogonal polynomials, which is evolved 
using the lattice Boltzmann algorithm.
The validation of our proposed scheme is performed in the context 
of the one-dimensional boost invariant expansion (Bjorken flow)
at various values of the ratio $\eta / s$ of the shear viscosity 
to the entropy density. This study is limited to the case of 
massless particles obeying Maxwell-J\"uttner statistics.
\end{abstract}

\section{INTRODUCTION}
\label{sec:intro}

Following the seminal work of Israel and Stewart \cite{israel}, 
relativistic viscous fluid dynamics has been continuously developed 
for a wide range of applications, including stellar collapse \cite{fryer04}, 
accretion problems \cite{banyuls97}, cosmology \cite{ellis12} and the 
quark-gluon plasma (QGP) \cite{jacak12} (see \cite{romatschke17} for a recent review).

The formation of the QGP was highlighted in the mid-rapidity range of 
ultra-high energy heavy ion collisions during the experiments 
performed at the Relativistic Heavy Ion Collider (RHIC) 
(for a historical account, please see Ref.~\cite{muller15}).
The experimental evidence validated the boost-invariant 
longitudinal expansion model proposed by Bjorken \cite{bjorken83},
which was initially solved in the perfect fluid approximation.

Even though the experimental evidence confirmed that the QGP exhibited 
an extremely low shear viscosity ($\eta$) to entropy density ($s$) ratio,
the time scale of its lifespan and the pressure anisotropy induced by the 
longitudinal expansion prevent the QGP from settling into complete 
thermodynamic equilibrium, as required for a perfect fluid. Indeed, 
through the adS/CFT conjecture, it was determined that 
$\eta / s = \hbar / 4\pi k_B$ for relativistic conformal
field theories at finite temperature and zero chemical potential
\cite{kovtun05}. In the conditions prevalent during the lifetime 
of the QGP, viscosities of this order of magnitude can induce significant 
deviations from the perfect fluid limit \cite{romatschke07}.

The construction of a relativistic viscous hydrodynamics theory 
has proven to be a formidable challenge, since the correspondent of 
the Navier-Stokes equations represents a first order formulation 
which is non-causal \cite{hiscock83}. The validation of second 
\cite{jaiswal13} and even third \cite{chattopadhyay15} order hydrodynamics 
equations relied on the semi-analytic solution of the relativistic Boltzmann 
equation for the one-dimensional boost-invariant longitudinal expansion 
\cite{florkowski13}. However, all hydrodynamic formulations 
seem to break down when $\eta / s$ becomes larger than some critical
value \cite{baier07}. Thus, it is clear that in order to perform 
realistic relativistic fluid dynamics simulations of QGP phenomena,
it is necessary to develop an efficient kinetic solver. This is the 
argument that motivates the present work.

The lattice Boltzmann method was 
recently employed to obtain solutions of the relativistic 
Boltzmann equation, including for the one-dimensional boost-invariant 
longitudinal expansion \cite{romatschke11,ambrus18qr}. The implementation 
in Ref.~\cite{ambrus18qr} was validated against the semi-analytic solution 
of the Bjorken flow presented in Ref.~\cite{florkowski13} for 
massless particles obeying Maxwell-J\"uttner statistics at 
vanishing chemical potential, but only for the case of isotropic 
initial conditions. Since it is expected that the strong longitudinal 
expansion induces a non-negligible pressure anisotropy, 
it is necessary that the scheme developed in Ref.~\cite{ambrus18qr} 
be extended to the case of anisotropic initial conditions, 
which we implement by using the Romatschke-Strickland 
form \cite{romatschke03}. This is the main result of this paper.
The resulting scheme is validated by comparison with first and 
second order hydrodynamics results at small $\eta / s$, analytic 
expressions in the free-streaming regime and the semi-analytic 
procedure derived in Ref.~\cite{florkowski13} everywhere else.

\section{RELATIVISTIC BOLTZMANN EQUATION FOR THE BJORKEN FLOW}
\label{sec:boltz}

The Bjorken flow is characterized by the assumption that the QGP 
properties in the mid-rapidity region (the central plateau) 
are invariant under Lorentz boosts along the longitudinal direction 
\cite{bjorken83}. This assumption fixes the
macroscopic four-velocity
at $u^t = t/\tau$ and $u^z = z / \tau$ while 
neglecting the transverse dynamics (i.e., $u^x = u^y = 0$),
where $\tau = \sqrt{t^2 - z^2}$ is the proper time.\footnote{The reference 
velocity is the speed of light $\widetilde{c}$ and the reference time 
is the initial proper time $\widetilde{\tau}_0$. All quantities 
bearing a tilde are dimensionful.}
The analysis of the flow properties is simplified by introducing the Milne coordinates 
$(\tau, x, y, \eta_M)$, where 
$\eta_M = \tanh^{-1} (z/t)$ is the space-time rapidity, with respect to which 
the Minkowski line element becomes 
$ds^2 = -d\tau^2 + dx^2 + dy^2 + \tau^2 d\eta_M^2$.
It is convenient to introduce the following tetrad vector frame $e_\halpha$ 
($e_\htau = \partial_\tau$, $e_{\hat{x}} = \partial_x$, $e_{\hat{y}} = \partial_y$
$e_{\heta_M} = \tau^{-1} \partial_{\eta_M}$) 
and its associated one-form coframe $\omega^\halpha$
such that $p^\halpha = \omega^\halpha_\mu p^\mu$ \cite{ambrus18qr}.

As a first approximation, the QGP can be effectively described as a parton 
gas dominated by (massless) gluons obeying Maxwell-J\"uttner statistics 
at vanishing chemical potential \cite{florkowski13,ambrus18qr}. 
The mass shell condition $\eta_{\halpha\hbeta} p^\halpha p^\hbeta = 0$ implies that 
$p^{\htau} = \sqrt{(p^\hatx)^2+(p^\haty)^2 + (p^{\heta_M})^2} \equiv p$. 
The non-dimensionalised Maxwell-J\"uttner equilibrium 
distribution $\feq$ at vanishing chemical potential can be written as:
\begin{equation}
 \feq = \frac{\widetilde{p}_{\rm ref}^3}{\widetilde{n}^{\rm(eq)}_0} 
 \widetilde{f}^{\rm (eq)} = \frac{1}{8\pi} e^{-p/T},
 \label{eq:feq}
\end{equation}
where $\widetilde{p}_{\rm ref} = \widetilde{k}_B \widetilde{T}_0 / \widetilde{c}$
is the reference momentum, $\widetilde{n}_0^{\rm(eq)} = 
8 \pi g_s (\widetilde{k}_B \widetilde{T}_0 / 2\pi \widetilde{\hbar} \widetilde{c})^3$ 
is the initial equilibrium parton number density at vanishing chemical potential 
and initial temperature $\widetilde{T}_0$, $g_s = 16$ represents the gluon number
of degrees of freedom, while the temperature
$T = P^{1/4}$ is given in terms of the isotropic pressure $P$ of the parton gas.

The evolution of the parton distribution function $f \equiv f(\tau; p, \xi)$ 
which depends on the momentum magnitude $p$ and $\xi = p^{\heta_M} / p^\htau$ 
is given by the relativistic Boltzmann equation written with respect to 
tetrad fields \cite{cardall13} as \cite{ambrus18qr}:
\begin{equation}
 \partial_\tau f + \frac{1}{\tau} f -
 \frac{\xi^2}{\tau p^2} \partial_p (fp^3) - \frac{1}{\tau} \partial_\xi[\xi(1-\xi^2) f]
 = -\frac{1}{\tau_{\rm{A-W}}} [f - \feq],\label{eq:boltz_Milne}
\end{equation}
where the gas is assumed to be homogeneous with respect to the space-time 
rapidity $\eta_M$ and transverse coordinates $x$ and $y$. The 
right hand side of the above equation represents the Anderson-Witting 
approximation of the Boltzmann collision integral and the 
relaxation time $\tau_{\rm A-W}$ is given by \cite{ambrus18qr}:
\begin{equation}
 \tau_{\rm A-W} = \frac{\tau_{\rm A-W;0}}{P^{1/4}}, \qquad 
 \tau_{\rm A-W;0} = \frac{5\hbar \etas}
 {\widetilde{\tau_0} \widetilde{k}_B \widetilde{T}_0} 
 \simeq 0.523426 \times 
 \frac{0.25\ {\rm fm}}{\widetilde{c} \widetilde{\tau}_0}\times
 \frac{600\ {\rm MeV}}{\widetilde{k}_B \widetilde{T}_0}
 \times 4\pi \etas.
\end{equation}
The above implementation of $\tau_{\rm A-W}$ ensures that 
the shear viscosity to entropy density ratio $\etas$ 
is constant for sufficiently small values of $\tau_{\rm A-W}$ 
\cite{florkowski13} (the overline denotes that the ratio 
is expressed in Planck units). The analysis in this
paper is restricted to the case when 
$\widetilde{c} \widetilde{\tau}_0 = 0.25\ {\rm fm}$
and $\widetilde{k}_B \widetilde{T}_0 = 600\ {\rm MeV}$.
The longitudinal and transverse pressures 
$\mathcal{P}_L$ and $\mathcal{P}_T$, the isotropic pressure 
$P$ and the pressure deviator $\Pi$ can be computed as moments of $f$:
\begin{equation}
 \mathcal{P}_L = \int \frac{d^3p}{p^\htau} f\, (p^{\heta_M})^2, \qquad 
 \mathcal{P}_T = \int \frac{d^3p}{p^\htau} f\, (p^\hatx)^2
 = \int \frac{d^3p}{p^\htau} f\, (p^\haty)^2, \qquad 
 P = \frac{1}{3}(\mathcal{P}_L + 2 \mathcal{P}_T), \qquad 
 \Pi = \frac{2}{3}(\mathcal{P}_L - \mathcal{P}_T).
 \label{eq:macro}
\end{equation}

%

In order to allow for an anisotropy of the distribution function at the 
initial time $\tau = \tau_0 = 1$, $f$ can be initialized using the 
Romatschke-Strickland form \cite{romatschke03}, which has 
the following dimensionful expression:
\begin{equation}
 \widetilde{f}_{\rm R-S} = \frac{g_s}{(2\pi \widetilde{\hbar})^3} 
 \exp\left[-\frac{1}{\widetilde{\Lambda}_0} 
 \sqrt{(\widetilde{p} \cdot \widetilde{u})^2 + \xi_0 \widetilde{c}^2 (\widetilde{p} \cdot z)^2}\right],
 \label{eq:frs_dim}
\end{equation}
where $\xi_0 > -1$ is a measure of the initial pressure anisotropy along the 
direction of the unit vector $z$ and $\widetilde{\Lambda}_0$ is an energy 
scale fixed by specifying the initial temperature.\footnote{The case $\xi < 0$ corresponding
to a prolate pressure anisotropy ($\mathcal{P}_L > \mathcal{P}_T$) 
is not considered in this paper.}
In the case of the Bjorken flow, $u^\halpha = (1,0,0,0)^T$
and $z^\halpha = (0,0,0,1)$ is the space-time rapidity unit vector, such that 
Eq.~\eqref{eq:frs_dim} can be expressed in non-dimensional form as:
\begin{equation}
 f_{\rm R-S} = \frac{1}{8\pi} \exp\left[-\frac{p}{\Lambda_0} \sqrt{1 + \xi_0 \xi^2}\right],
 \qquad 
 \Lambda_0 = \left[\frac{1}{2} 
 \left(\frac{1}{\sqrt{\xi_0}} \arctan \sqrt{\xi_0} + 
 \frac{1}{1 + \xi_0}\right)\right]^{-1/4},\label{eq:frs}
\end{equation}
valid for any $\xi_0 > 0$. The initial longitudinal $\mathcal{P}_{L,0}$ 
and transverse $\mathcal{P}_{T,0}$ pressures are:
\begin{equation}
 \mathcal{P}_{L,0} = 
 \frac{3\Lambda_0^4}{2\xi_0} \left[ \frac{1}{\sqrt{\xi_0}} \arctan\sqrt{\xi_0} - 
 \frac{1}{1 + \xi_0}\right],\qquad
 \mathcal{P}_{T,0} = \frac{3\Lambda_0^4}{4\xi_0} 
 \left(1 + \frac{\xi_0 - 1}{\sqrt{\xi_0}} \arctan\sqrt{\xi_0}\right),
\end{equation}
while $P_0 = 1$. The 
initial pressure deviator $\Pi_0$ is:
\begin{equation}
 \Pi_0 = -\frac{\Lambda_0^4}{2\xi_0}\left[\frac{\xi_0 + 3}{\xi_0 + 1} + 
 \frac{1}{\sqrt{\xi_0}} (\xi_0 - 3) \arctan \sqrt{\xi_0}\right] 
 = -\left[1 - \frac{3}{\xi_0} + \frac{6}
 {\xi_0 + \sqrt{\xi_0} (\xi_0 + 1) \arctan \sqrt{\xi_0}}\right].
 \label{eq:Pi0}
\end{equation}

\section{LATTICE BOLTZMANN ALGORITHM}

In order to solve Eq.~\eqref{eq:boltz_Milne}, we employ the lattice Boltzmann 
algorithm introduced in Ref.~\cite{ambrus18qr}, which we extend in order 
to account for the anisotropic initial conditions given by Eq.~\eqref{eq:frs}.
The momentum space is discretized following the prescriptions 
of Gauss quadratures with respect to the spherical coordinates
$(p, \theta, \varphi)$. Since the flow is isotropic in the transverse 
plane, the $\varphi$ degree of freedom is integrated analytically.
The magnitude $p$ of the momentum is discretised using the generalized Gauss-Laguerre 
quadrature of order $Q_L = 2$ [$L_{Q_L}^{(2)}(p_k) = 0$, $1 \le k \le Q_L$], 
while the Gauss-Legendre quadrature of order $Q_\xi$ is used to discretise 
$\xi = \cos\theta$ [$P_{Q_\xi}(\xi_j) = 0$, $1 \le j \le Q_\xi$]. 
The above discretization allows $\mathcal{P}_L$ and $\mathcal{P}_T$ 
\eqref{eq:macro} to be computed using the following quadrature sums \cite{ambrus18qr}:
\begin{equation}
 \mathcal{P}_T = \frac{1}{2} \sum_{j = 1}^{Q_{\xi}} \sum_{k= 1}^{Q_L} f_{jk} p_k (1 - \xi_j^2), \qquad 
 \mathcal{P}_L = \sum_{j = 1}^{Q_{\xi}} \sum_{k= 1}^{Q_L} f_{jk} p_k \xi_j^2,
 \label{eq:moments}
\end{equation}
where $f_{jk} = 2\pi e^{p_k} w_j^\xi w_k^L f(p_k, \xi_j)$, while 
$w_k^L$ and $w_j^\xi$ are the Gauss-Laguerre and Gauss-Legendre quadrature weights.

In order to ensure conservation of energy-momentum, 
the equilibrium distribution 
$\feq$ appearing on the right hand side of Eq.~\eqref{eq:boltz_Milne} 
is replaced by a truncated series with respect to the Laguerre and 
Legendre polynomials.
In the particular case of the Bjorken flow, $\feq$ does not depend on $\xi$, 
such that its expansion is simply:
\begin{equation}
 \feq_{jk} = \frac{1}{4} n^{\rm(eq)} w_j w_k [1 + (1 - T)(3 - p_k)], \qquad 
 n^{\rm(eq)} = P^{3/4}, \qquad T = P^{1/4},
\end{equation}
while $P = \frac{1}{3}(2\mathcal{P}_T + \mathcal{P}_L)$ is obtained 
using the moments \eqref{eq:moments} of $f$.
Starting from the expansion of $f$ with respect to the Laguerre and 
Legendre polynomials, the derivatives with respect to $p$ and $\xi$ appearing in 
Eq.~\eqref{eq:boltz_Milne} can be written as
$\left[p^{-2} \partial_p(fp^3)\right]_{jk} = 
\sum_{k' = 1}^{Q_L} \mathcal{K}^L_{k,k'} f_{j,k'}$ 
and $\left\{\partial_\xi[\xi(1-\xi^2) f]\right\}_{jk} = 
\sum_{j' = 1}^{Q_\xi} \mathcal{K}^P_{j,j'} f_{j',k}$, where 
the kernels $\mathcal{K}^L_{k,k'}$
$\mathcal{K}^P_{j,j'}$ are given in Ref.~\cite{ambrus18qr}.
The time stepping in Eq.~\eqref{eq:boltz_Milne} is 
performed using a third order Runge-Kutta 
algorithm \cite{shu88} with $\delta \tau = 10^{-3}$.

%

\begin{figure}
\begin{tabular}{ccc}
\includegraphics[angle=0,width=0.32\linewidth]{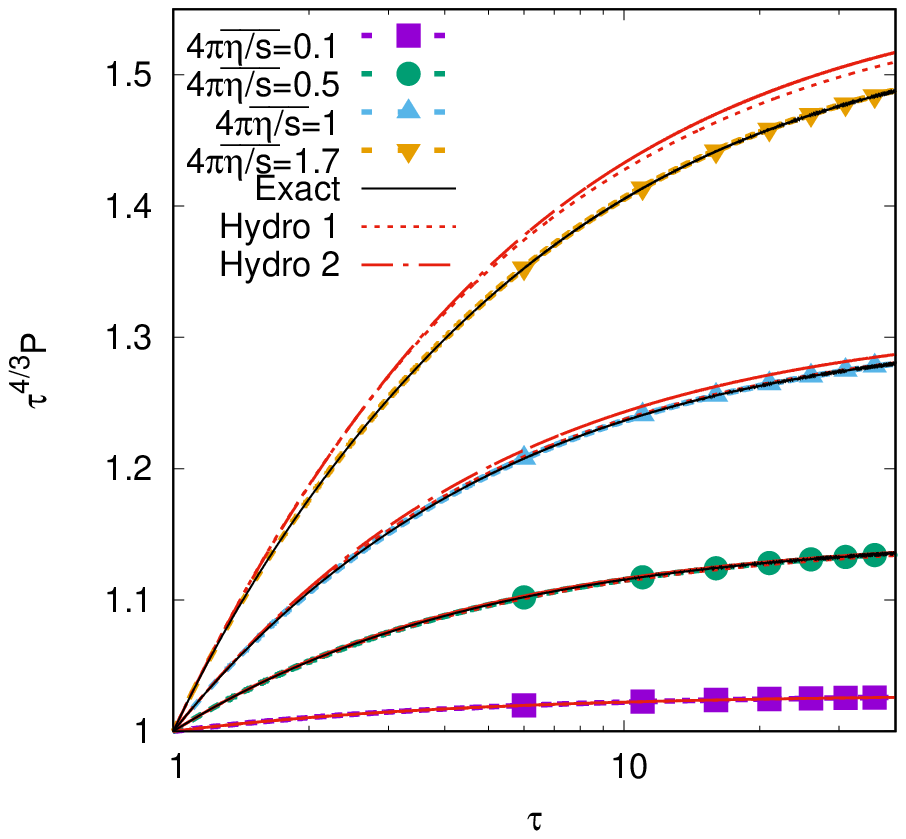} &
\hspace{-10pt}
\includegraphics[angle=0,width=0.32\linewidth]{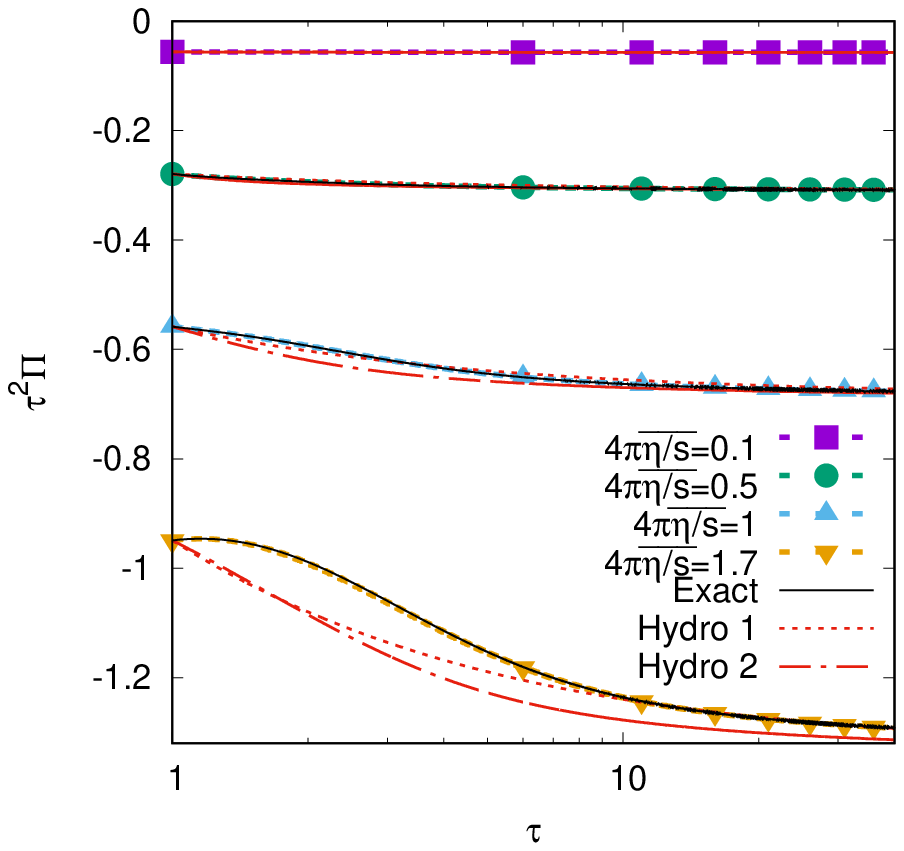} &
\hspace{-10pt}
\includegraphics[angle=0,width=0.32\linewidth]{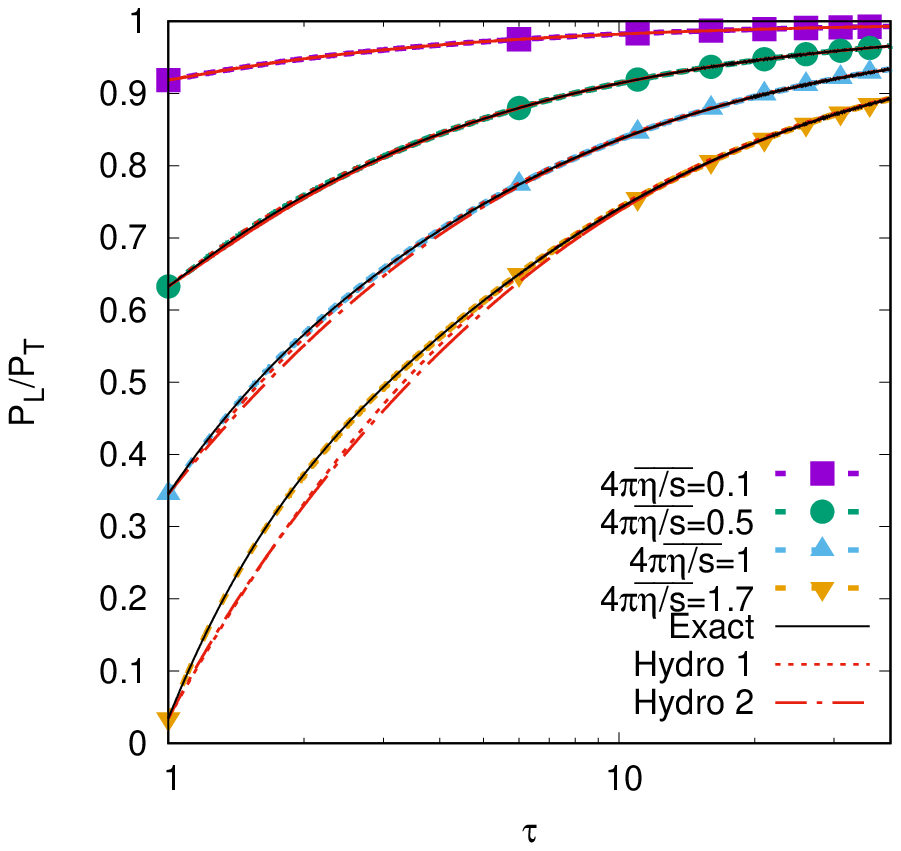} 
\end{tabular}
\caption{Comparison between LB results (lines and points), first-order (dotted lines) 
and second-order (dash-dotted lines) hydrodynamics and
the semi-analytic exact solution reported in Ref.~\cite{florkowski13}
[shown using solid lines only for $4\pi \etas >0.1$],
for $P$ (left), $\Pi$ (center) and 
$\mathcal{P}_L / \mathcal{P}_T$ (right)
for various values of $\etas$.
The anisotropy parameter $\xi_0$ is found by 
solving Eq.~\eqref{eq:ximagic} and has the values 
$\xi_0^{\rm H1} \simeq 0.11$ [$\etas = 0.1$], 
$0.76$ [$\etas = 0.5$], $2.67$ [$\etas = 1$] and 
$48.4$ [$\etas = 1.7$].\label{fig:ximagic}}
\end{figure}

The expansion of $f_{\rm R-S}$ \eqref{eq:frs} 
with respect to the generalized Laguerre polynomials is:
\begin{equation}
 f_{\rm R-S} = e^{-p} \sum_{\ell = 0}^\infty \frac{1}{(\ell + 1) (\ell + 2)} 
 \mathcal{F}_{{\rm R-S}; \ell} L_\ell^{(2)}(p), \qquad 
 \mathcal{F}_{{\rm R-S}; \ell} = \int_0^\infty dp\, p^2\, f_{\rm R-S} L_{\ell}^{(2)}(p).
\end{equation}
Due to the orthogonality of the Laguerre polynomials, only the coefficients 
corresponding to $\ell = 0$ and $\ell = 1$ contribute to the 
dynamics of $T^{\halpha\hbeta}$ \cite{ambrus18qr}. Substituting 
$L_0^{(2)}(p) = 1$ and $L_1^{(1)}(p) = 3 - p$, the following 
expressions are found:
\begin{equation}
 \mathcal{F}_{{\rm R-S};0} = \frac{\Lambda_0^3}{4\pi (1 + \xi_0 \xi^2)^{3/2}}, \qquad 
 3 \mathcal{F}_{{\rm R-S};0} - \mathcal{F}_{{\rm R-S}; 1} = 
 \frac{3\Lambda_0^4}{4\pi (1 + \xi_0 \xi^2)^2}.
\end{equation}
It remains to expand the functions $(1 + \xi_0 \xi^2)^{-n/2}$ with respect to the 
Legendre polynomials:
\begin{equation}
 \frac{1}{(1 + \xi_0 \xi^2)^{n/2}} = \sum_{m = 0}^\infty \frac{2m+1}{2} a^{\rm R-S}_{n,m} 
 P_m(\xi), \qquad 
 a^{\rm R-S}_{n,m} = \int_{-1}^1 \frac{d\xi\, P_m(\xi)}{(1 + \xi_0 \xi^2)^{n/2}},
\end{equation}
such that $f_{\rm R-S}$ is replaced by:
\begin{equation}
 f_{{\rm R-S};jk} = \frac{\Lambda_0^3}{4} w_j^{\xi} w_k^{L} \sum_{m = 0}^{N_\Omega} 
 \frac{2m + 1}{2} P_m(\xi_j) \left[a^{\rm R-S}_{3,m} +
 (3 - p_k) (a^{\rm R-S}_{3,m} - \Lambda_0 a^{\rm R-S}_{4,m}) \right],
\end{equation}
where $0 \le N_\Omega < Q_\xi$ represents the truncation order of $f_{\rm R-S}$.
In general, increasing $N_\Omega$ also increases the accuracy of the 
simulations and the sensitivity to $N_\Omega$ increases with $\xi_0$. 
The results reported in this paper were obtained with $N_\Omega = 6$.
Noting that $a^{\rm R-S}_{n,m}$ with odd $m$ vanish, the following 
coefficients were used:
\begin{eqnarray}
 a^{\rm R-S}_{3,0} = \frac{2}{\sqrt{1 + \xi_0}}, \quad 
 a^{\rm R-S}_{3, 2} = \frac{3 {\rm arcsinh}\sqrt{\xi_0}}{\xi_0^{3/2}} - 
 \frac{3 + \xi_0}{\xi_0 \sqrt{1 + \xi_0}}, \quad 
 a^{\rm R-S}_{3,4} = \frac{105 + 95\xi_0 + 6\xi_0^2}{8\xi_0^2\sqrt{1 + \xi_0}} - 
 \frac{15(7 + 4\xi_0) {\rm arcsinh}\sqrt{\xi_0}}{8 \xi_0^{5/2}}, \nonumber\\
 a^{\rm R-S}_{3,6} = \frac{105(33 + 36\xi_0 + 8 \xi_0^2) {\rm arcsinh} \sqrt{\xi_0}}
 {64 \xi_0^{7/2}} - 
 \frac{3465 + 4935 \xi_0 + 1638 \xi_0^2 + 40 \xi_0^3}
 {64 \xi_0^3 \sqrt{1 + \xi_0}}, \qquad
 a^{\rm R-S}_{4,0} = \frac{1}{1 + \xi_0} + \frac{\arctan\sqrt{\xi_0}}{\sqrt{\xi_0}}, \nonumber\\
 a^{\rm R-S}_{4,2} = \frac{1}{1 + \xi_0} - \frac{3}{2\xi_0} + 
 \frac{3 - \xi_0}{2\xi_0^{3/2}} \arctan \sqrt{\xi_0}, \quad
 a^{\rm R-S}_{4,4} = \frac{105 + 100\xi_0 + 3\xi_0^2}{8\xi_0^2 (1 + \xi_0)} - 
 \frac{3(35 + 10\xi_0 - \xi_0^2)}{8 \xi_0^{5/2}} \arctan\sqrt{\xi_0},\quad\nonumber\\
 a^{\rm R-S}_{4,6} = \frac{5(231 + 189\xi_0 + 21\xi_0^2 - \xi_0^3)}
 {16\xi_0^{7/2}} \arctan{\sqrt{\xi_0}} - 
 \frac{1155 + 1715\xi_0 + 581 \xi_0^2 + 5\xi_0^3}{16\xi_0^3(1 + \xi_0)}.\qquad
 \qquad\qquad
\end{eqnarray}

\section{NUMERICAL ANALYSIS}

\begin{figure}
\begin{tabular}{ccc}
\includegraphics[angle=0,width=0.32\linewidth]{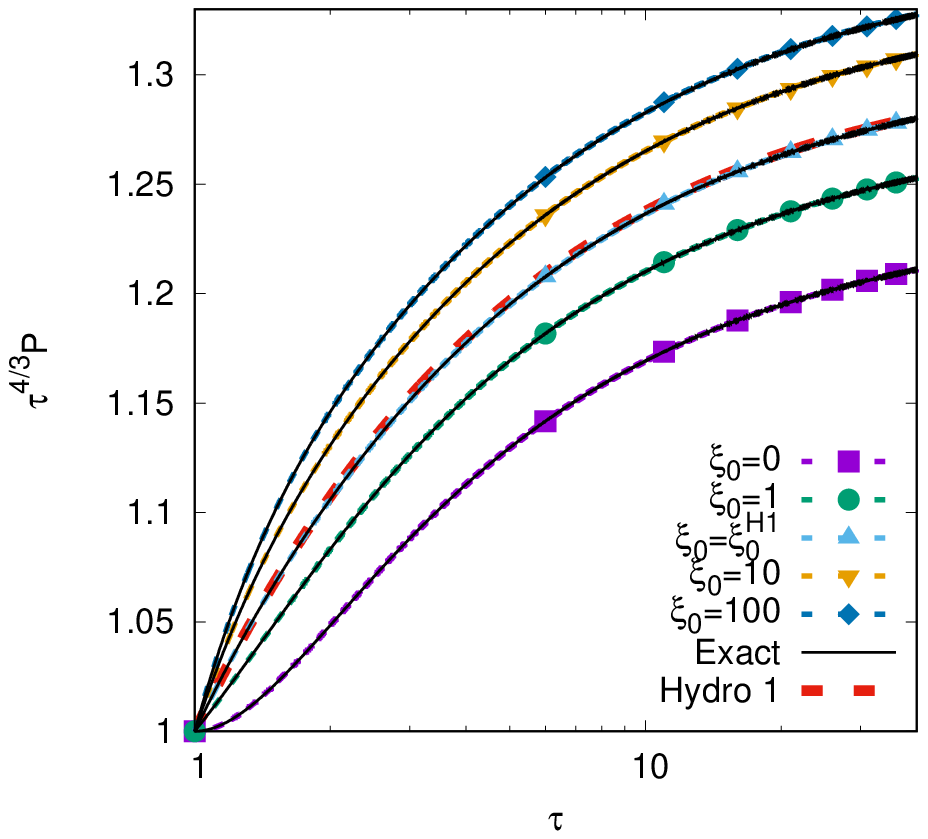} &
\hspace{-10pt}
\includegraphics[angle=0,width=0.32\linewidth]{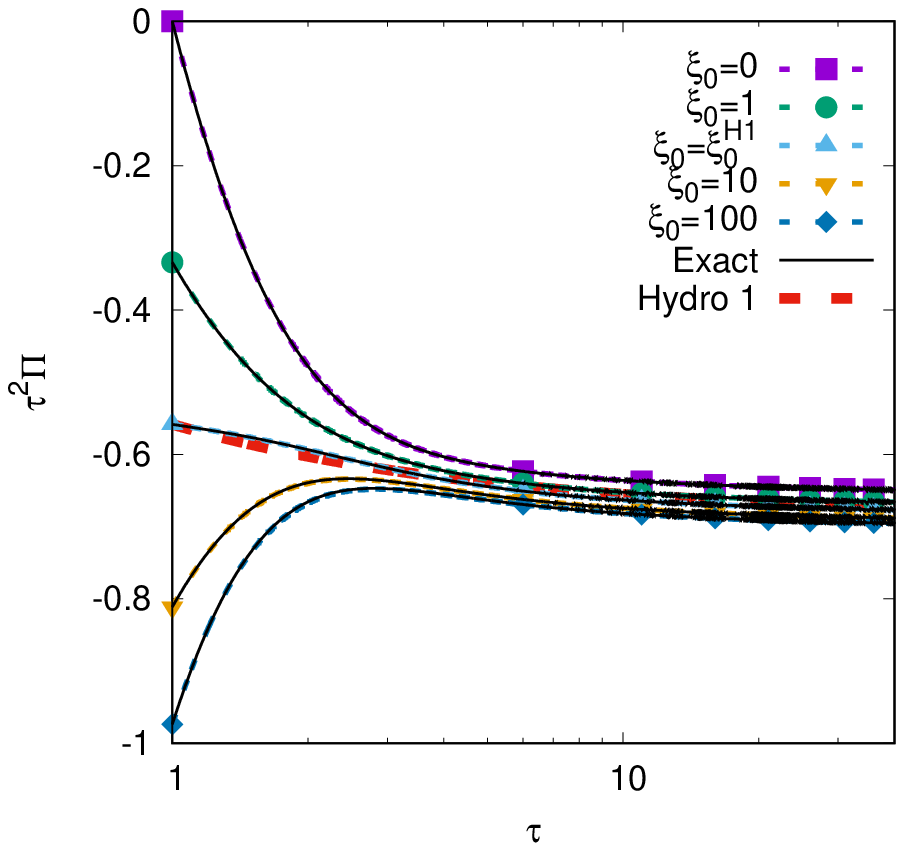} &
\hspace{-10pt}
\includegraphics[angle=0,width=0.32\linewidth]{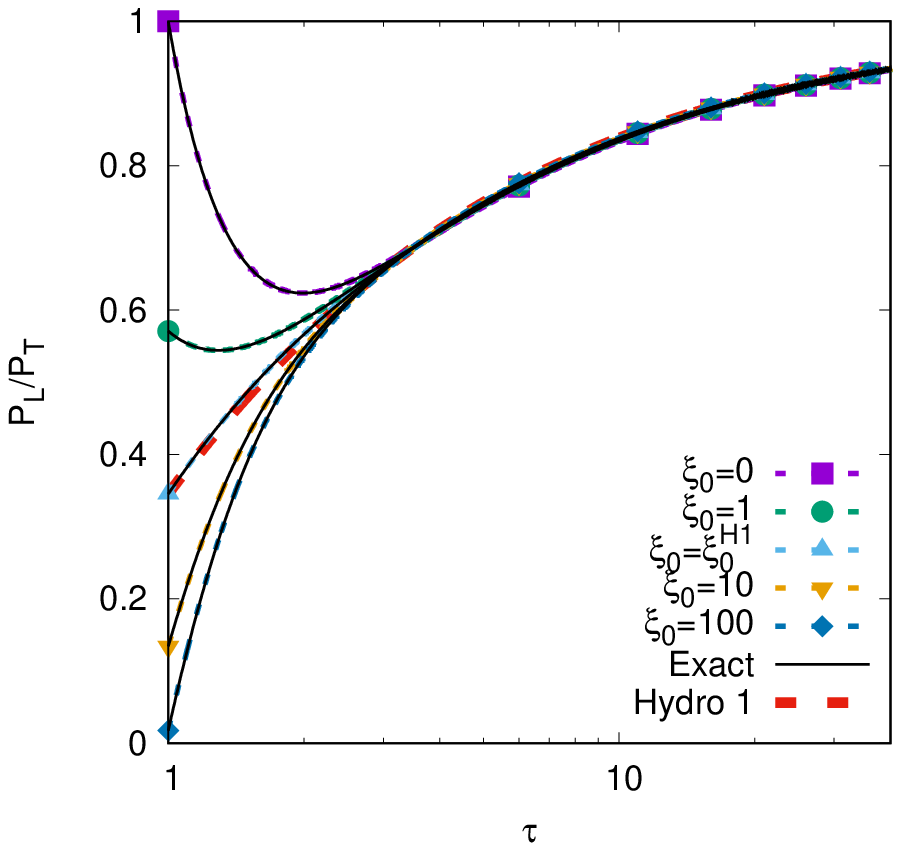}
\end{tabular}
\caption{Comparison between the LB results obtained using $Q_\xi = 20$ 
and the exact semi-analytic solution reported in Ref.~\cite{florkowski13}
for $\xi_0 \in \{0, 1, \xi_0^{\rm H1}, 100\}$, 
where $\xi_0^{\rm H1}\simeq 2.67$ is obtained from Eq.~\eqref{eq:ximagic},
at $4\pi \etas = 1$.
\label{fig:etas1}}
\end{figure}

The conservation of the stress-energy tensor $\nabla_\hbeta T^{\halpha\hbeta}$ reduces 
for the Bjorken flow to $3\tau \partial_\tau P + 4P + \Pi = 0$.
In the first-order hydrodynamics theory, 
$\Pi = -\frac{4\eta}{3\tau} = 
-\frac{16 \tau_{\rm A-W;0}}{15\tau} P^{3/4}$
where $\eta = \frac{4}{5} \tau_{\rm A-W} P$ is the 
Chapman-Enskog value for the shear viscosity \cite{cercignani02}.
In the second-order hydrodynamics theory, the constitutive equation for 
$\Pi$ is promoted to an evolution equation \cite{jaiswal13,ambrus18qr}.
It can be seen that in the first order theory, $\Pi_0 = -\frac{16}{15} \tau_{\rm A-W;0}$,
which can only be achieved via the Romatschke-Strickland form \eqref{eq:frs}
when $\tau_{\rm A-W;0} < \frac{15}{16}$.\footnote{ 
At larger values of 
$\tau_{\rm A-W;0}$, the initial longitudinal pressure 
$\mathcal{P}_{L,0} = P_0 + \Pi_0 = 1-\frac{16}{15} \tau_{\rm A-W;0}$ 
would attain negative values, which is forbidden in the parton
gas model considered in this paper, indicating the breakdown 
of the first order hydrodynamics description.}
The value $\xi_0^{\rm H1}$ corresponding to this value
is the solution of the following equation 
($\widetilde{k}_B \widetilde{T}_0 = 600\ {\rm MeV}$,
$\widetilde{c} \widetilde{\tau}_0 = 0.25\ {\rm fm}$):
\begin{equation}
 1 - \frac{3}{\xi_0^{\rm H1}} + \frac{6} 
 {\xi_0^{\rm H1} + \sqrt{\xi_0^{\rm H1}} 
 (\xi_0^{\rm H1} + 1) \arctan \sqrt{\xi_0^{\rm H1}}} 
 \simeq 0.558321 \times 4\pi \etas.
 \label{eq:ximagic}
\end{equation}
It is thus natural to consider a comparison between the first- and 
second-order hydrodynamics for the case when $\Pi_0$ coincides in the 
two theories (i.e., $\xi_0 = \xi_0^{\rm H1}$).
Figure~\ref{fig:ximagic} shows that, quite remarkably, 
the first-order theory seems to be in better agreement with the 
numerical results obtained using the LB algorithm presented in the 
previous section for $4\pi\etas \in \{0.1, 0.5, 1, 1.7\}$. It is particularly 
interesting to note that, at $4\pi \etas = 1.7$, the late time evolution 
of $\Pi$ overlaps with the first-order prediction, while the 
second-order hydrodynamics result remains in visible disagreement 
even at $\tau = 40$.

Next, Fig.~\ref{fig:etas1} shows the effect of varying the initial anisotropy 
$\xi_0$ at $4\pi \etas = 1$. It can be seen that at fixed proper time $\tau$, the 
pressure increases as $\xi_0$ is increased. The first order hydrodynamic 
prediction seems to be a late time hydrodynamic attractor for $\Pi$ and 
more strikingly for $\mathcal{P}_L / \mathcal{P}_T$.

\begin{figure}
\begin{tabular}{ccc}
\includegraphics[angle=0,width=0.32\linewidth]{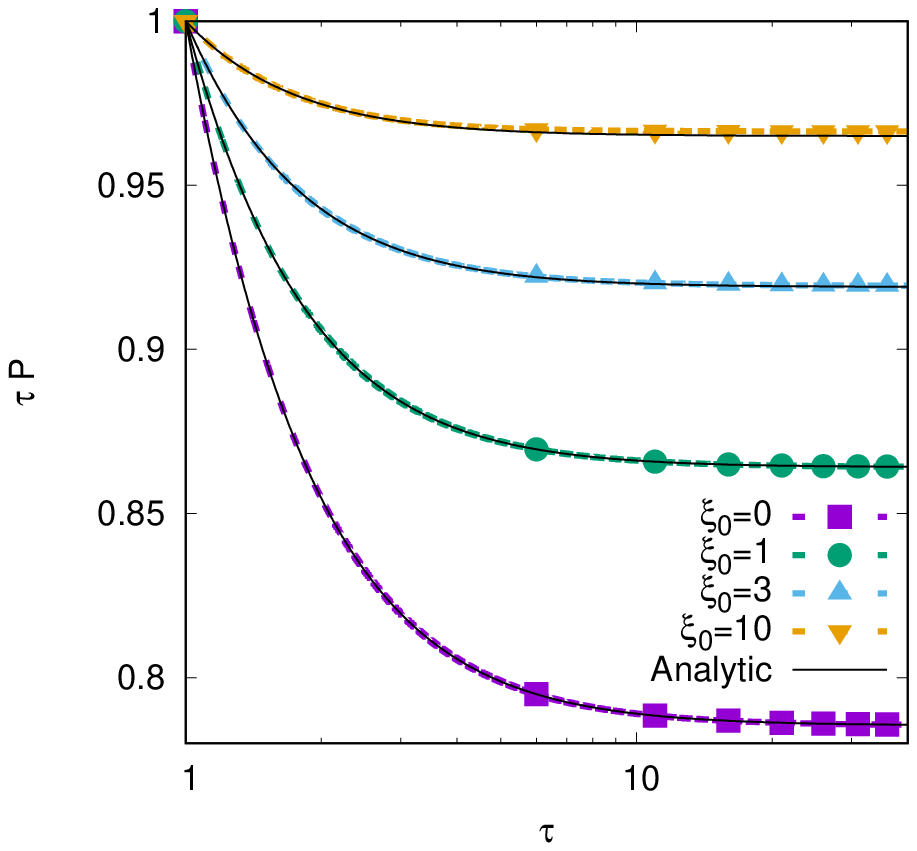} &
\hspace{-10pt}
\includegraphics[angle=0,width=0.32\linewidth]{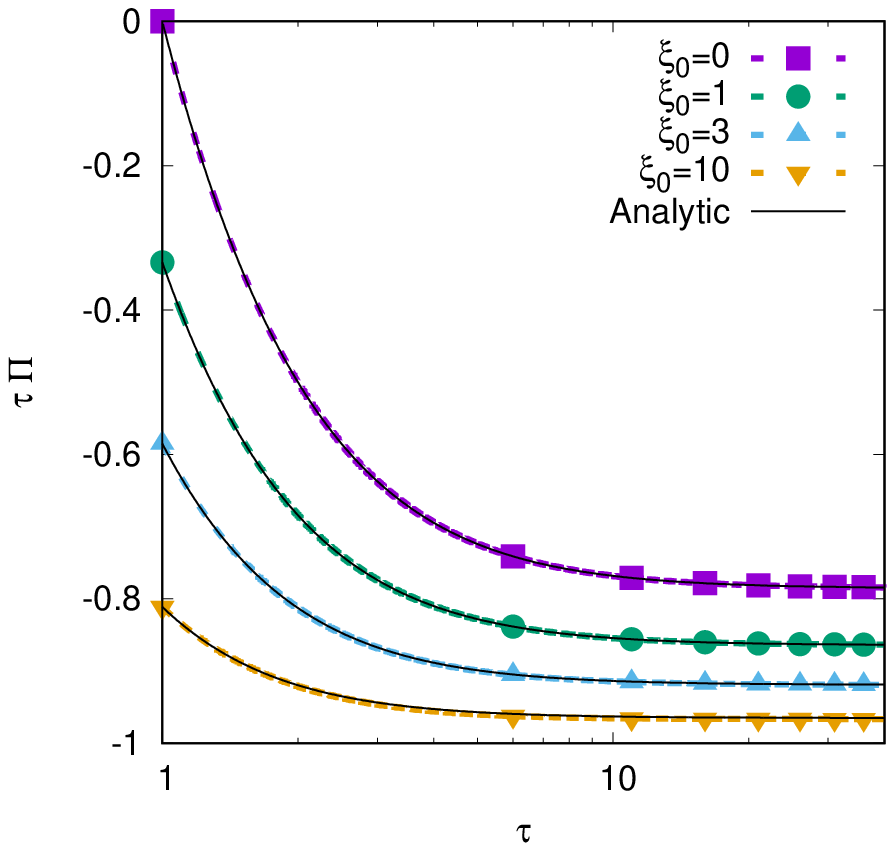} &
\hspace{-10pt}
\includegraphics[angle=0,width=0.32\linewidth]{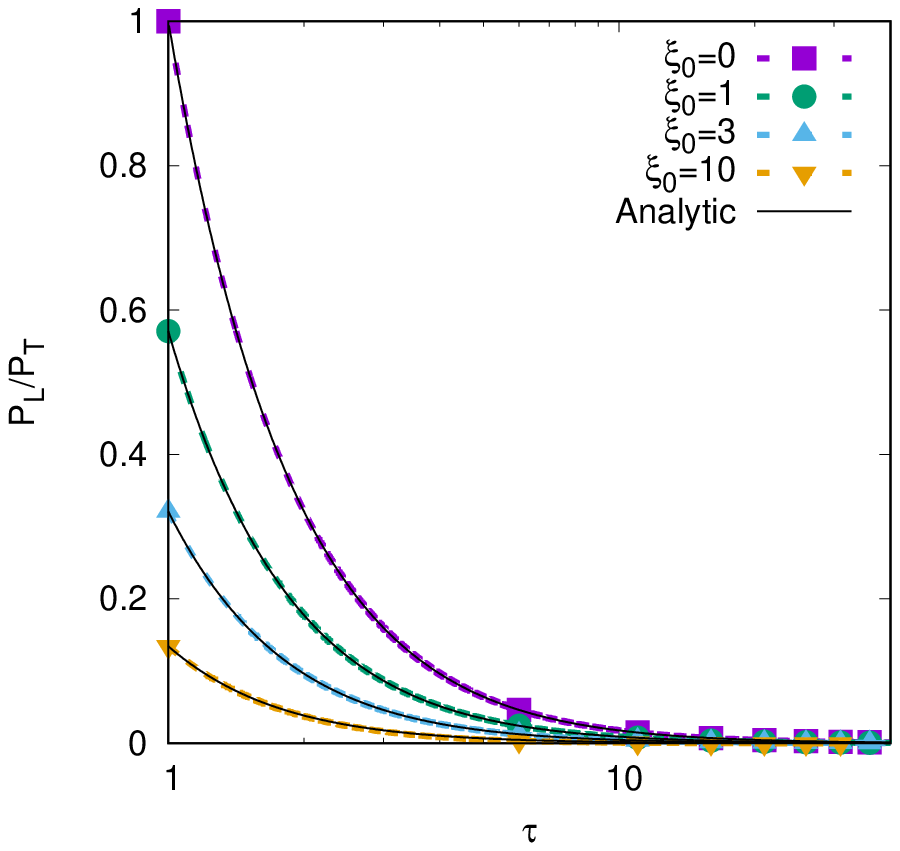} 
\end{tabular}
\caption{Comparison between LB results (lines and points)
for the free-streaming case and the analytic solution 
\eqref{eq:bal} for various values of $\xi_0$.\label{fig:bal}}
\end{figure}

In the free-streaming limit, $\tau_{\rm A-W;0} \rightarrow \infty$ and 
the right-hand side of Eq.~\eqref{eq:boltz_Milne} vanishes, such that:
\begin{equation}
 f = \frac{1}{8\pi} \exp\left\{-\frac{p}{\Lambda_0} 
 \sqrt{1 + \xi^2[(1 + \xi_0)\tau^2 - 1]}\right\},
\end{equation}
which corresponds to the following moments:
\begin{eqnarray}
 \mathcal{P}_{L,{\rm fs}} &=&
 \frac{3\Lambda_0^4}{2[(1+\xi_0)\tau^2 - 1]} \left[ \frac{\arctan{\sqrt{(1+\xi_0)\tau^2-1}}}
 {\sqrt{(1+\xi_0)\tau^2-1}} - 
 \frac{1}{(1 + \xi_0)\tau^2}\right],\nonumber\\
 \mathcal{P}_{T,{\rm fs}} &=& 
 \frac{3\Lambda_0^4}{4[(1+\xi_0)\tau^2 - 1]} \left\{
 1 + [(1+\xi_0)\tau^2-2] \frac{\arctan{\sqrt{(1+\xi_0)\tau^2-1}}}
 {\sqrt{(1+\xi_0)\tau^2-1}}\right\}.
 \label{eq:bal}
\end{eqnarray}
It can be seen in Fig.~\ref{fig:bal} that the numerical results obtained using 
$Q_\xi = 40$ and $N_\Omega = 6$ are in very good agreement with the above analytic 
results. The small discrepancy observed at $\xi_0 = 10$ becomes unnoticeable
when $N_\Omega = 8$.

\section{CONCLUSIONS}

In this paper, we developed a procedure to implement anisotropic 
initial conditions for the one-dimensional boost-invariant longitudinal 
expansion (Bjorken flow) in the frame of the lattice Boltzmann (LB) algorithm 
for the relativistic Boltzmann equation introduced in Ref.~\cite{ambrus18qr}. 
The accuracy of our implementation was demonstrated by comparison with the 
solutions of the first- and second-order hydrodynamics equations, with 
the analytic solution of the free streaming limit and with the 
semi-analytic solution derived in Ref.~\cite{florkowski13}. In all cases, 
an excellent agreement was found. The implementation of the initial anisotropy 
via an expansion of order $N_\Omega = 6$ with respect to the Legendre polynomials 
of the Romatschke-Strickland form of the Maxwell-J\"uttner distribution
gave accurate results for $0 \le \xi_0 \le 100$ at finite values of $\etas$, 
however small deviations form the exact solution could be seen 
in the free-streaming limit for $\xi=10$. The accuracy of the simulations 
in this case can be improved by increasing $N_\Omega$.

When $\xi_0$ is used to set the initial value of the pressure deviator $\Pi_0$ 
to the value required by the first-order constitutive equations, our simulations
show that the first-order theory remains in very good agreement with the 
numerical results even at $4\pi \etas = 1.7$, where the second-order 
formulation is no longer accurate. 

Increasing $\xi_0$ at fixed $4\pi \etas = 1$, the initial pressure anisotropy 
can be directly controlled, 
with $\mathcal{P}_L / \mathcal{P}_T \rightarrow 0$ as $\xi_0 \rightarrow \infty$. 
For all tested values of $\xi_0$, we found that 
$\mathcal{P}_L / \mathcal{P}_T$ converges to the same curve for $\tau \gtrsim 5$. 
In the future, we plan to analyse this result and compare with 
the hydrodynamic attractor solution proposed in Ref.~\cite{romatschke17jhep}.
Interestingly, $P$ (and hence $T = P^{1/4}$) seems to increase 
monotonically with $\xi_0$ when $\tau$ is kept fixed.

In the future, we plan to extend our LB algorithm to account for 
massive particles \cite{gabbana17} and for particles obeying quantum 
(Fermi-Dirac and Bose-Einstein) statistics \cite{coelho18cf}. In the context 
of the Bjorken flow, our implementation can be validated against the semi-analytic 
solutions developed in Refs.~\cite{florkowski14} and \cite{florkowski15}, respectively. 

{\bf ACKNOWLEDGMENTS.}
This work was supported by a grant of the Romanian Ministry of Research and Innovation,
CCCDI-UEFISCDI, project number PN-III-P1-1.2-PCCDI-2017-0371, within PNCDI III.


\begin{thebibliography}{99}

\bibitem{israel}
W. Israel, {\it Ann. Phys.} {\bf 100}, 310 (1976); 
W. Israel and J. M. Stewart, {\it Ann. Phys.} {\bf 118}, 341 (1979).

\bibitem{fryer04}
C. L. Fryer, {\it Stellar Collapse} (Kluwer Academic Publishers, Dordrecht, Netherlands, 2004).

\bibitem{banyuls97}
F. Banyuls, J. A. Font, J. M. Ibanez, J. M. Marti, and J. A. Miralles, 
{\it Astrophys. J.} {\bf 476}, 221231 (1997).

\bibitem{ellis12}
G. F. R. Ellis, R. Maartens, and M. A. H. MacCallum, {\it Relativistic cosmology}
(Cambridge University Press, Cambridge, UK, 2012).

\bibitem{jacak12}
B. V. Jacak and B. Muller, {\it Science} {\bf 337}, 310314 (2012).

\bibitem{romatschke17}
P. Romatschke and U. Romatschke, arXiv:1712.05815 [nucl-th].

\bibitem{muller15}
B. M\"uller, ``A New Phase of Matter: Quark-Gluon Plasma
Beyond the Hagedorn Critical Temperature'', in 
{\it Melting hadrons, boiling quarks}, edited by J. Rafelski 
(Springer, 2015), DOI: 10.1007/978-3-319-17545-4, pp. 107--116.

\bibitem{bjorken83}
J. D. Bjorken, {\it Phys. Rev. D} {\bf 27}, 140--151 (1983).

\bibitem{kovtun05}
P. Kovtun, D. T. Son, and A. O. Starinets, {\it Phys. Rev. Lett.}
{\bf 94}, 111601 (2005).

\bibitem{romatschke07}
P. Romatschke and U. Romatschke, {\it Phys. Rev. Lett.} 
{\bf 99}, 172301 (2007).

\bibitem{hiscock83}
W. A. Hiscock and L. Lindblom, {\it Ann. Phys.} {\bf 151}, 466--496 (1983).

\bibitem{jaiswal13}
A. Jaiswal, {\it Phys. Rev. C} {\bf 87}, 051901 (2013).

\bibitem{chattopadhyay15}
C. Chattopadhyay, A. Jaiswal, S. Pal, and R. Ryblewski,
{\it Phys. Rev. C} {\bf 91}, 024917 (2015).

\bibitem{florkowski13}
W. Florkowski, R. Ryblewski, and M. Strickland,
{\it Phys. Rev. C} {\bf 88}, 024903 (2013).

\bibitem{baier07}
R. Baier and P. Romatschke, {\it Eur. Phys. J. C} {\bf 51}, 677--687 (2007).

\bibitem{romatschke11}
P. Romatschke, M. Mendoza, and S. Succi, 
{\it Phys. Rev. C} {\bf 84}, 034903 (2011).

\bibitem{ambrus18qr}
V. E. Ambru\cb{s} and R. Blaga, {\it Phys. Rev. C} {\bf 98}, 035201 (2018).

\bibitem{romatschke03}
P. Romatschke and M. Strickland, {\it Phys. Rev. D} {\bf 68}, 
036004 (2003).

\bibitem{cardall13}
C. Y. Cardall, E. Endeve, and A. Mezzacappa, 
{\it Phys. Rev. D} {\bf 88}, 023011 (2013).

\bibitem{shu88}
C.-W. Shu and S. Osher, {\it J. Comput. Phys.} {\bf 77}, 439--471 (1988).

\bibitem{cercignani02}
C. Cercignani and G. M. Kremer, 
{\it The relativistic {B}oltzmann equation: theory and applications}
(Birkh\"{a}user Verlag, Basel, Switzerland, 2002).

\bibitem{romatschke17jhep}
P. Romatschke, {\it J. High Energy Phys.} {\bf 12} (2017) 079.

\bibitem{gabbana17}
A. Gabbana, M. Mendoza, S. Succi, and R. Tripiccione, 
{\it Phys. Rev. E} {\bf 95}, 053304 (2017).

\bibitem{coelho18cf}
R. C. V. Coelho, M. Mendoza, M. M. Doria, and H. J. Herrmann, 
{\it Comput. Fluids} {\bf 172}, 318--331 (2018).

\bibitem{florkowski14}
W. Florkowski, E. Maksymiuk, R. Ryblewski, and M. Strickland,
{\it Phys. Rev. C} {\bf 89}, 054908 (2014).

\bibitem{florkowski15}
W. Florkowski and E. Maksymiuk, 
{\it J. Phys. G: Nucl. Part. Phys.} {\bf 42}, 045106 (2015).


\end{thebibliography}
\end{document}